# Optimal Daily Trading of Battery Operations Using Arbitrage Spreads

**Ekaterina Abramova *,† and Derek Bunn †**

Department of Management Science and Operations, London Business School, Regent's Park, London NW1 4SA, UK; dbunn@london.edu
* Correspondence: eabramova@london.edu
† These authors contributed equally to this work.

**Abstract:** An important revenue stream for electric battery operators is often arbitraging the hourly price spreads in the day-ahead auction. The optimal approach to this is challenging if risk is a consideration as this requires the estimation of density functions. Since the hourly prices are not normal and not independent, creating spread densities from the difference of separately estimated price densities is generally intractable. Thus, forecasts of all intraday hourly spreads were directly specified as an upper triangular matrix containing densities. The model was a flexible four-parameter distribution used to produce dynamic parameter estimates conditional upon exogenous factors, most importantly wind, solar and the day-ahead demand forecasts. These forecasts supported the optimal daily scheduling of a storage facility, operating on single and multiple cycles per day. The optimization is innovative in its use of spread trades rather than hourly prices, which this paper argues, is more attractive in reducing risk. In contrast to the conventional approach of trading the daily peak and trough, multiple trades are found to be profitable and opportunistic depending upon the weather forecasts.

## 1. Introduction

Operating a storage facility within an electricity system used to be relatively straightforward. Pumped hydro was, until a few years ago, the only technology available at sufficient scale and would typically operate on a daily cycle: pumping during the off-peak periods at night and generating during the peak day and evening periods. In a wholesale market context, the intraday hourly price swings were generally predictable and sufficiently large to ensure a regular daily return on the price arbitrage between the off-peak pumping costs and on-peak generation revenues. However, the incursion of renewable generating technologies, wind and solar, has fundamentally changed the intraday price dynamics so much that depending upon the weather, the size and timing of potential arbitrage spreads vary considerably day-by-day. In sunny locations with substantial solar energy, e.g., California, the lowest hourly prices may often be in the middle of the day, since widespread local solar production reduces demand on the wholesale market [1]. Thus, the expected intraday spreads in hourly prices, and their consequent arbitrage opportunities for load shifting or storage, will depend upon solar and wind forecasts, as well as demand and supply considerations. Operating a storage facility has therefore become riskier and more dependent upon accurate, weather-dependent, hourly price forecasts [2].

Furthermore, the introduction of intermittent generation by wind and solar has increased the value of batteries. Batteries are more economic, flexible resources than pumped hydro and can be introduced at small as well as large scale. Batteries are often co-located with solar or wind facilities to avoid curtailment [3] or included in asset portfolios by aggregators who may also be trading demand-side management services [4]. They may have additional revenue sources from providing the increased real-time reserve or balancing services required by the grid to deal with the greater volatility in power flows due to the intermittent technologies [5,6]. Although valuations of merchant battery assets take account of multiple revenue streams [7], seeking a profitable arbitrage position from day-ahead (DA) auctions remains a crucially important ingredient [8]. The focus of this paper is therefore on the optimal day-ahead, risk constrained arbitrage positions (where 'risk' refers to the loss of money due to roundtrip transaction costs and battery inefficiency upon charge/discharge, i.e., only the downside and not upside risk is considered) leading to daily cycles for the battery operations.

There are many attractive reasons for battery operators to evaluate the arbitrage spreads based upon the day-ahead market prices. The day-ahead auctions for wholesale power typically close before midday and set prices simultaneously for all 24 h in the next day. This means that buy-sell pairs can be acquired at the same time without leaving any open positions, and that operational planning can be done for the following day. Furthermore, these auctions are more liquid than the continuous intraday trading that follows, and as uniform auctions, they do not incur the larger bid-ask spreads often seen in the continuously trading order books. For example, the most liquid intraday trading in Europe is in Germany, but the bid-ask spreads there have varied from approx. €15 several hours ahead to about €3 for a few minutes ahead of delivery, and they are larger elsewhere in Europe [9]. In Britain, there has been low liquidity intraday until an hour before delivery [10]. Therefore, the acquisition of spread pairs in the day-ahead auctions is convenient, less risky and of lower cost than using the continuous order books on the intraday exchanges. Furthermore, with power market dynamics typically showing daily price cycles with large intraday price swings, daily optimization, rather than over longer cycles, offers more revenue cycles and tends to be the norm in theory and practice. An aversion to the self-discharge losses in batteries holding charges for several days may also motivate shorter operational cycles [11]. Finally, the microstructure of trading power means that high granularity products (hourly, half-hourly or less) are only available day-ahead and that trading further forward is in longer products such as daily baseload (24 h) or peak (12 or 14 h daytime). So, planning daily cycles and fixing the profits day-ahead is attractive, and locking in the profit from the day-ahead auction is financially efficient. However, the auction presents a challenge in that the trader needs to decide which buy-sell pair(s) they want to acquire and target their bids and offers accordingly. This requires precise forecasting and risk analysis. That is the methodological starting point for this paper.

Operational examples of daily scheduling include Foster [12] who reports on the daily cycling of a large Tesla battery in S. Australia, Mohsenian-Rad [13] who computes optimal battery schedules from the day-ahead market of California, Salles et al. [14] who undertake nodal valuations of batteries based upon day-ahead arbitrages in PJM, and Lucas [4] who documents the actual daily operational cycle of a commercial battery in London. Therefore, there is widespread evidence that profit maximization based upon daily cyclical operations, ideally using day-ahead auction prices for the hourly spreads, remains an important revenue stream for battery operators, even if supplemented with other revenues for services to the grid operators and retailers. Furthermore, even if the operator is considering other modes of operation, the authors of this paper have been informed by a battery operator that the estimation of potential arbitrage profits from the day-ahead auction can provide an opportunity cost baseline for bidding the asset into other flexibility auctions, as offered by grid operators, or for some of the storage services required by end-users and retailers. However, despite these profit maximization algorithms being based upon expected arbitrage spreads in the hourly prices, as set by day-ahead auctions, risk is an important consideration, as emphasized by Lucas [4]. The "roundtrip" costs of a battery cycle include efficiency losses, degradation, trading and use of system costs, and hence operators typically look for a safe expected margin above these before committing to an arbitrage trade. This is a cautious pragmatic response to risk aversion and it is surprising that despite its prevalence in risk management, there has not been a "value-at-risk" approach to this storage trading risk, e.g., in terms of a 95% chance of exceeding the roundtrip costs, and it would seem natural to seek to apply this standard and a more precise formulation. However, there are methodological challenges in doing so and therein lies the contribution of this paper.

The formulation of battery operation in this paper is based upon acquiring profitable intraday price spreads from the day-ahead auctions. For a trader to decide whether and how much to bid for purchase, or offer to sell power at specific hours in the auction, forecasts of the price spreads are needed before the auction closes. Furthermore, since a probabilistic model based upon value-at-risk (VaR) is sought after in this paper, density forecasts of the spreads are required. In addition, it is possible that more than one spread trade could be feasible if the forecast profile of the hourly prices is sufficiently volatile. This is where the methodology developed in this paper is innovative. Although the day-ahead electricity price forecasting has been a topic of substantial and broad research in terms of methods, the focus has mostly been on price levels for the delivery periods (usually hourly) in the following day. Hence, based upon day-ahead forecasts of the drivers of electricity prices such as demand, gas and coal prices, wind and solar production, forecasts for electricity price levels have been proposed from various methods [15–17]. In the last few years, there has been an interest in density forecasts for the hourly prices, driven by risk management considerations [18,19].

Since the 24-hourly day-ahead prices emerge simultaneously from the auction, each hour is not independent of the other hours. Furthermore, the price density functions for each hour are non-normal with skewness switching between negative and positive depending upon the dynamics of production of renewable energy [20]. So, creating the density function of the difference of two price densities would be analytically elusive. Hence the necessity to forecast the spread densities for each pair of hourly arbitrage possibilities separately. The functional form of these densities must be sufficiently flexible and responsive to the weather drivers. Not only does wind and solar production influence the average spreads, but they also change the higher moments, variance, skewness and kurtosis of the density functions, which are necessary for accurate value-at-risk estimates. So, it is necessary to have spread densities which have a functional relationship to weather forecasts, available day-ahead.

To achieve this, the generalized additive model for location, scale and shape (GAMLSS) parametric regression model is used [21], which has previously been successfully applied to obtain day-ahead densities of price levels [20] and price spreads [22]. This paper extends that latter work on the econometrics of spreads to provide the basis for optimal battery operations. Within the framework of this paper, the hourly electricity price spreads form response variables, whose distribution functions vary according to multiple exogenous factors. Thus, the daily spread densities in prices, and their consequent value-at-risk constrained arbitrage opportunities, will depend upon wind and solar forecasts, as well as demand forecasts and other fundamentals. The dynamic location, scale and shape parameters (dynamic “latent moments” connected to the mean, volatility, skewness and kurtosis of price spreads) are thus directly incorporated into the forecasting model. An alternative approach to the specification of density functions would be to use quantile regression, as this has been increasingly used to forecast the percentiles of the distribution function based on exogenous factors [15,23]. However, the attraction of an analytically tractable density function is that the computation of specific tail probabilities can be derived directly from the density specification itself, rather than being approximated by interpolation between several quantile estimates.

This paper applies the spread densities formulation to the German market, which forms the largest and main daily reference for wholesale power in Europe. This market is also strongly influenced by wind and solar production, as well as providing a place where batteries and demand-side management are active innovations. In fact, this paper tests a conjecture that on days with high wind and/or solar, creating a midday trough in demand on the wholesale market, optimal operation may be more than one cycle per day.

The next sections review some of the background research on optimal battery operations followed by the formulation of an optimization model. The data are then described, and the features of the forecast spread densities, which are taken from Abramova et al. [22] are summarized. In the subsequent section the results of using these density functions to create the optimal daily scheduling of battery storage are reported and the superior profitability compared to using normal densities is demonstrated through backtesting. The final section offers concluding comments.

## 2. Background Research

This paper is positioned distinctively within the large body of research on energy storage, at the interface of operations and finance, as well as within sustainable supply chains where batteries play an increasingly important role following the emergence of renewable energy resources. The increased research and industrial innovation in electricity storage is therefore strongly linked to the growth of renewable energy. In this wider context Hu et al. [24] and Aflaki et al. [25] consider the intermittency effect on renewable investment. Wu et al. [26] consider the renewable curtailment effects of intermittency while Qi et al. [3] evaluate the co-location of storage. Regarding price formation Al-Gwaiz et al. [27] consider the renewable effects on supply function competition, Kim et al. [28] formulate optimal electricity forward contracts with storage, while Zhou et al. [29] look at the impact of negative prices on optimal electricity storage decision rules. The use of electric vehicles is a special kind of battery operation within the grid, and Broneske et al. [30] and Kahlen et al. [31] consider optimal contracting and vehicle-to-grid operations. White et al. investigate electricity grid arbitrage using seven re-purposed EV batteries [32], showing the extended value of EV batteries in energy storage applications. Beyond electricity Wu et al. [33] contrast the daily cycles for electricity with the annual cycles for natural gas storage. They focus upon the longer-term gas storage as a dynamic real options optimization, with a focus on the futures prices. Secomandi [34] also focuses upon gas storage as a dynamic inventory problem and in Nadarajah et al. [35] extends this to deal with network transportation trading. However, in contrast to a large body of work on gas and other storable commodities, e.g., [36,37], because of the predominance of the day-ahead auctions and the daily periodicity in electricity prices, the operational horizon for storage and load shifting is typically episodic daily. As emphasized above, this paper does not suggest daily cycling based upon arbitraging the day-ahead prices is the singular way to optimize battery operations, rather it is one revenue stream in a possible portfolio that battery operators will consider.

Within power systems research, the dominant formulation for optimizing battery operations has been over the day-ahead horizon. In part this reflects the well-established daily unit commitment scheduling by system operators [38], as well as the cycling of pumped storage which has a long history of daily operations [39]. Furthermore, the major role of the day-ahead auction is to provide a liquid and efficient market to fix the prices for day-ahead contracted volumes. Stochastic optimization is increasingly being applied within this setting, since, even if price risk is removed by day-ahead trading, output volume remains uncertain, especially with intermittent resources. Among the many published day-ahead stochastic optimization formulations, the following are indicative of different research directions taken. None, however, have explicitly used spread trading to determine the operations.

Market power is one possible issue and the impact of a large price-moving storage unit in the DA market has been studied [40]. A bi-level problem has been modeled for this purpose. The upper-level problem comprises of profit maximization of the storage unit, while market clearing is considered a lower-level problem. Similarly, the profit maximization problem for a geographically dispersed and large price-maker battery participating in a nodal electricity market has been modeled [41]. Several factors including the size, location, charge and discharge rates and efficiency of the energy storage units, as well as the influence of the energy storage operations on the locational marginal prices, were considered. Looking at operational dynamics, the effect of ramping inflexibility of conven-tional generators on the operation of a price making merchant energy storage has been studied [42]. In all of these, the price that the energy storage receives is calculated in the lower-level problem by the market-clearing algorithm that maximizes the social welfare.

The interaction of day-ahead and subsequent real-time operations has been considered by several researchers. In Krishnamurthy et al. [43] a comparison was performed between the cases when the storage owner only submits quantity bids, and when they submit both price and quantity bids in the DA market. An ARIMA point forecast was used together with a quantile-copula kernel density estimator for probabilistic forecasts. Hybrid facilities and multiple income streams are also a major topic of research. Thus, in Tohidi et al. [44] the stochastic optimization comprises of flow battery storage, photovoltaic (PV) plant and load. Two revenue streams that are investigated include (i) arbitrage between DA market and balancing market, (ii) during the peak load hours supplying the local PV production through the storage. Similarly, a stochastic valuation of grid-scale energy storage in the wholesale markets considering several revenue streams simultaneously, has been performed [45]. A coordinated operation of energy storage and a wind farm in day-ahead and intraday electricity markets, inspired by the Spanish electricity market, has been considered [46].

While focusing upon determining day-ahead operations, a few researchers have observed that it may be useful to anticipate operations more than a day-ahead. Thus, a look-ahead technique taking into account both DA and the following day has been proposed in Wang et al. [47] to decide optimal bidding strategy for a merchant energy storage operator. The consideration of discounted profit opportunities helps in optimizing the state of charge at the end of the target day. The optimization problem is a stochastic bi-level problem. In Aliasghari et al. [48], the usefulness of look-ahead to operate an integrated system comprising of conventional generators, wind power and compressed air energy storage is evaluated.

While the day-ahead planning lends naturally to stochastic optimization over the short term daily or longer horizons, higher frequency, continuous trading of a battery asset is often modeled using SDP. Trading a battery based on real-time or close to real-time scenario is a different perspective. It is opportunistic and more speculative. Unlike the day-ahead spreads, it may involve open positions, formulated as optimal states of charge for the battery at particular moments in time regarding expectations on future price movements. In Jiang et al. [49], for example, the problem of hour-ahead optimal bidding of an energy storage operator in the real-time electricity market has been modeled using ADP. The operator faces the challenge to bid in the market without having the knowledge about the battery level at the start of the hour and at the same time considering the value of remaining energy at the end of the hour. This issue has been addressed by formulating an algorithm that can exploit the monotonicity of the value function to be able to obtain the revenue-generating bidding policy. Similarly, Shu et al. [50] propose an adaptive policy for an energy storage in a grid-connected wind power plant, in which wind power output and electricity prices are the two uncertainties considered in the revenue model. With risk considerations, a multistage problem with the aim to minimize quantile-based risk measure for the future costs at each stage was proposed for the bidding problem of a risk-averse energy storage [51]. The stochasticity of electricity prices, wind power output and load are important aspects of all these high frequency adaptive models. In Durante et al. [52], the stochastic prices and wind power forecast errors have been modeled by a hidden semi-Markov model. This model can accurately capture the error distribution as well as the crossing times to represent when each process is above or below a certain benchmark, for example the forecast. This makes the process a partially observable MDP, which is then solved using a backward ADP approach. The sequential decision problem to find the optimal control policies of an energy storage can be modeled using an SDP approach. However, with a large state space, the traditional techniques become computationally intractable. In [53], several ADP algorithms have been used to solve such a finite horizon energy storage problem, to be able to reach the near-optimal policies with reduced computational burdens taking account of the stochastic wind supply and electricity prices.

Looking at the co-optimization problem of a storage device that can be used for multiple purposes, for example energy arbitrage, ancillary services, distribution relief, and backup energy, Xi et al. [54] used an approximate version of SDP in two-phases. The decision variables and exogenous state are discretized, and the objective function values are found using DSDP approach, in the first phase. In the second phase, these objective function values from the first phase are used for piecewise linearization of the value function of the true SDP, thus forming a mixed integer program which is then solved for a near-optimal policy. Finally, electric vehicles present a similar operational challenge and Jiang et al. [55] model the behavior of the electric vehicle charging station manager, who attempts to meet the charging demand while minimizing the cost of charging. This problem has been modeled as a finite horizon MDP with a dynamic risk measure.

Thus, there are many formulations for optimal operations of battery assets and each focuses upon specific aspects. The formulation used in this paper, takes a day-ahead stochastic optimization approach, which has several distinctive, under-researched characteristics. It focuses upon monetizing the battery assets through daily spread trading, which solves the problem of correlations between hourly prices when risk is being considered. Here the density functions for each spread are formulated so that risk constrained arbitrage can be specified. These density functions capture the dynamic switches in skewness, created by wind and solar output, using day-ahead weather forecasts. These are all novel features. Essentially, the decision problem this research envisages is that of a risk-averse battery operator, seeking to determine a provisional operating plan for a battery asset to know how to make bids and offers into the day-ahead auction. The outcome of the auction would fix the price and quantities for charging and discharging in the following day, therefore locking in the profit. This may be more attractive to risk-averse practitioners than continuously holding open positions on the state of charge of the battery. In practice, of course, a day-ahead schedule may be adjusted continuously through the day as opportunities arise. However, first, an optimal day-ahead schedule needs to be formulated, and that is the contribution of this research. The novelty of this formulation is in evaluating pure spread trading. This has some advantages, but also some constraints.

## 3. Stylized Properties of Spread Trades

The search for optimal daily spreads can be simplified by several observations on the nature of the periodic daily price process. Here the principles of fully balanced spread trading are examined, i.e., buying and selling equivalent volumes in the day-ahead market at different times, adjusted only for technical efficiency losses. This research does not optimize a mixture of spread trading and open positions. However, its analysis can be applied to a tranche of the battery exogenously reserved for fully balanced spread trading, with the remaining tranches reserved for other monetization. As indicated above, the spread traded operations have many attractions, but also several logical properties and constraints. For the initial perspective consider how the daily price cycle is characterized by several distinct turning points which can be usually quite accurately forecast depending upon weather and load forecasts. Figure 1 shows a typical daily profile which is taken from the data subsequently used for testing of the proposed trading model (data obtained for test time step $t = 5$).

Based upon simple considerations of this daily pattern, the following properties are observed:

**Property 1.** *Between two consecutive turning points in the price curve for which the price gradient is monotonic, it will never be optimal to trade intermediate price spread trades in the presence of transaction costs.*

**Proof of Property 1.** Let $i, j$ denote consecutive trough and peak times, $i < j$, so that $Y^{(i,j)}$ is the spread trade from trough to peak. Assume a non-zero transaction cost, $c$ and let $i < k < j$. Then $Y^{(i,j)}$ can be split into two consecutive spread trades $Y^{(i,j)} = Y^{(i,k)} + Y^{(k,j)}$. However, this split will cost $2c$ for two trades, compared to $c$ for the single trade. □

**Lemma 1.** *The maximum number of optimal spread trades over a finite horizon is the number of turning points (peaks and troughs) in the price function plus 2 for the end points.*

**Proof of Lemma 1.** This follows from P1, which demonstrates the exclusion of intermediate spread trades between turning points. □

**Property 2.** *Multiple trades determined from the day-ahead auction will either be sequential or embedded, they will not overlap each other. In other words, once the best single spread trade has been identified, the second-best trade will either completely precede or follow it, or will be an interior trade within its spread.*

**Proof of Property 2.** Consider a spread trade between a trough at time $i$ and a peak at time $j$ which is the best single trade in the day. Evidently there can exist another spread trade with a trough at time $k$ and a peak at time $l$ if $k < l < i$ or $j < k < l$. Such a $k, l$ spread trade would be a sequential second-best trade. Evidently an interior embedded second-best trade can exist if $i < k < l < j$. However, the following cannot exist: $i < k < j < l$, since it would imply a sequence of turning points $Min, Min, Max, Max$. For a smooth function, the turning points $Min, Max$ must alternate. So, a second-best spread trade cannot straddle the best trade. □

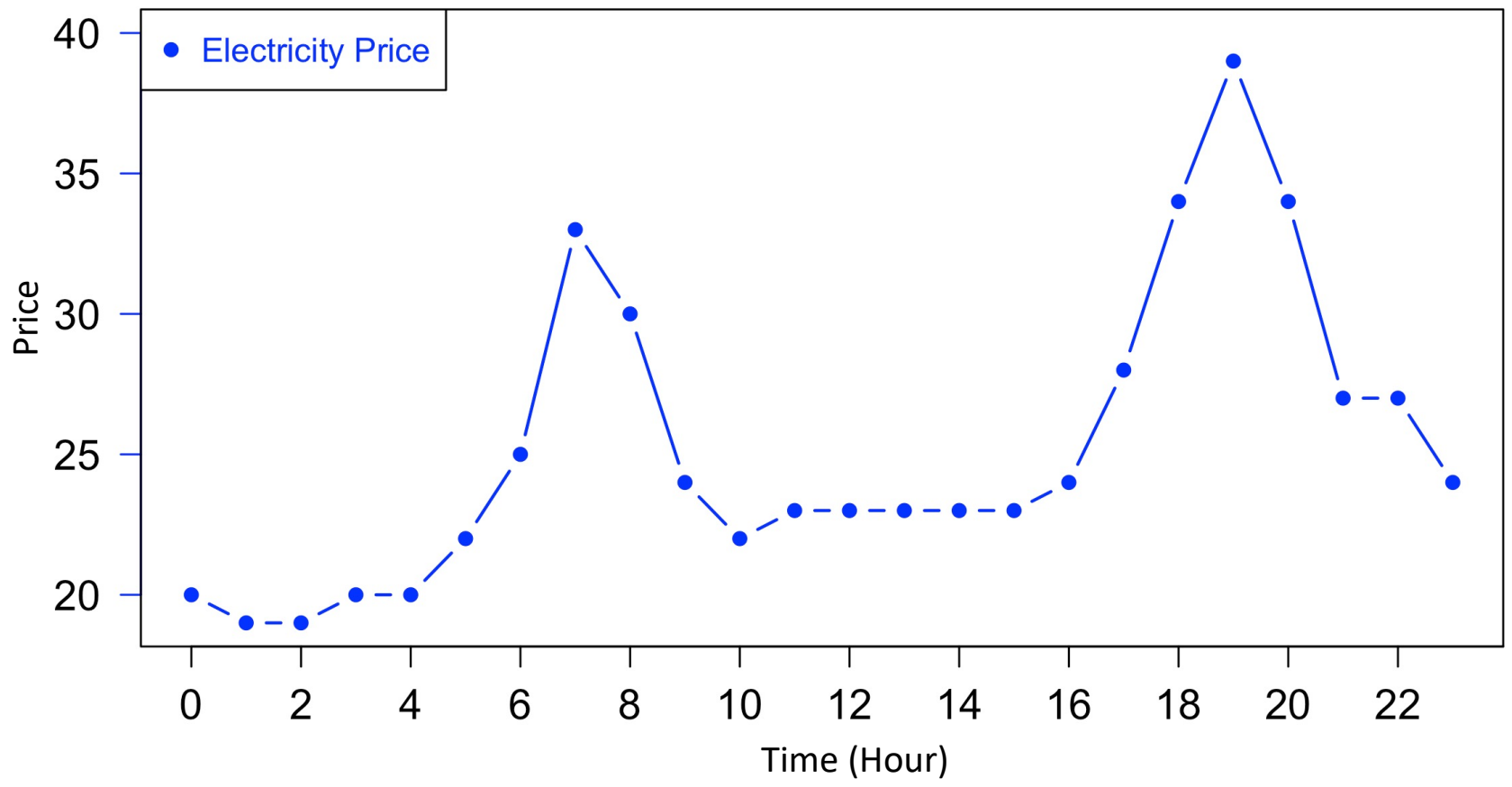

**Figure 1.** Typical German electricity price profile.

The implications of the above are that the search for optimal spreads can be restricted to consideration of the forecast turning points in the day-ahead price profile. Additionally, they can be identified sequentially, as the best, second best and so on. Furthermore, since the day-ahead auction allows prices and quantities to be fixed as outcomes of the single auction, i.e., the trade is closed, and alternative additional revenue streams from holding back capacity for intraday trading or network services are not considered, the following is observed:

**Property 3.** *With a fast charger that can charge or discharge in an hour as required, given a finite spread and a fixed transaction cost, the optimal plan will always be to charge or discharge to the maximum in the day-ahead spread trades.*

**Proof of Property 3.** This follows from P2 and the separability of multiple trades. Given that the spread trades are sequential or embedded, it will be optimal to charge or discharge at the maximum levels. □

**Lemma 2.** *The State of Charge (SoC) of the battery at any time will be binary zero or one.*

Lemma 2 follows directly from Proof of Property 3. Note: in practice it may not be efficient for a battery to fully discharge, but the limits $\{0, 1\}$ are used without loss of generality to indicate the practical limits of operation.

This has strong implications, achieving computational simplicity and may be surprising. However, it is intuitive on reflection by considering the stylized formulation involving transaction costs, fast charging or discharging within an hour, and the sequential nature of turning points with the forecast horizon. Furthermore, the optimal plan is being determined day-ahead based on the forecast day-ahead auction outcomes. It is not a sequential process of adjusting battery states of charge according to the option values created from a continuous stochastic price process. That is quite a different formulation, as discussed previously. Furthermore, the following is observed:

**Property 4.** *With only spread trading during the day, the opening and closing SoC will be the same.*

**Proof of Property 4.** Let $SoC_s$ and $SoC_f$ be the starting and finishing States of Charge. With one turning point in the day at time $i$, and using P3, the second spread trade $Y^{(i,f)}$ will completely reverse the first $Y^{(s,i)}$. With an extra turning point at $j$, following P2, without loss of generality it can be assumed $j > i$, and again $Y^{(j,f)}$ will completely reverse $Y^{(i,j)}$. And so on for further turning points, each spread fully reversing the previous one. □

**Lemma 3.** *The starting SoC for a continuous series of daily spread trading will lock-in all daily starting and finishing to the same value.*

Lemma 3 follows directly from Proof of Property 4.

This lemma provides further computational implications in that the state of charge at the beginning of each and every day in a continuous sequence of spread trading are the same and have effectively been set at the start of the sequence. Furthermore, the value will be either fully charged or fully discharged. This also implies that:

**Lemma 4.** *With fully balanced spread trading on the day-ahead auction the optimal operations are restricted to the following day.*

**Proof of Lemma 4.** Optimization over a longer horizon than one day would only be beneficial if the intervening $SoC_f$ could change, which, for spread trading, it cannot by P4. □

Evidently, if it is desirable to plan operations over longer time scales than the day-ahead, and trade accordingly, open positions will be required from the day-ahead auction, and by construction they would not be fully balanced spread trades. It does raise the question of whether spread trading is too restrictive, despite its other attractions, and whether more continuous, longer horizon strategies may be beneficial, despite their higher risks. This discussion will be resumed later. Next, development of the operational optimization model is continued with the fully balanced spread trading based on the simplifying properties established above.

## 4. Optimization

The starting point is the day-ahead forecasting of all pairwise price spreads. The day-ahead auction reveals separate hourly prices for the 24 h of next day, thus there is an upper triangular matrix of spread prices as auction outcomes. The random variable associated with each spread price between hours $i$ and $j$ is represented with $Y^{(i,j)}$. The matrix of these spread prices can also be thought of as a vector of $s = 1, \ldots, 276$ possible spreads (thus the simplified notation of $Y^{(s)}$ is also used where needed). Density forecasts were required to be made for each day-ahead spread in advance of the auction to evaluate the target spreads and their trading risks. The forecasting of these densities will be discussed in the next section. The forecasts are conditional upon weather and other data available each day to market participants ahead of the auction. For each of the 276 possible spread trades, the density function will give expectations as well as value-at-risk quantiles, e.g., $q_{05}$ and $q_{95}$ for the 5% and 95% points on the forecast spread densities. It is assumed that the objective function is to select one or more spread trades per day and to maximize the expected net payoff, subject to a value-at-risk constraint that the net payoff risk of each selected spread battery operation being less than the transaction costs is less than 5% . Based upon the forecasts, it is assumed the trader will make a competitive bid and offer into the auction to acquire the target trading pair(s).

Although the conceptual simplifications for fully balanced spread trading, as established in the previous section, would imply an integer formulation that can be readily solved by heuristic search, the daily optimization can be formally presented in the following way:

The profit contribution from spread trading is maximized per day as

$$\max_{\mathsf{b}_i,\mathsf{s}_j} \sum_{i=0}^{N-1}\sum_{j=0}^{N-1}\left[\frac{\eta}{2}Y^{(i,j)}(\mathsf{bs}_{ij}-1)(\mathsf{b}_i+\mathsf{s}_j)\right]-n*c \quad (1)$$

where $i, j$ are the hourly indices, $0\ldots23$, corresponding to the buy / sell trades respectively, $N$ is the number of hours in a day available for the battery to trade, $N = 24$, $n$ is the number of spread trades per day, $\eta$ is the battery efficiency for roundtrip spread trade (chemical efficiency of power conversion in a single storing and discharging cycle), $c$ is the transaction costs for a roundtrip spread trade, $\mathsf{b}_i$ is the amount of energy (non-negative) to be bought by the battery at hour $i$ (long position), $\mathsf{s}_j$is the amount of energy (non-negative) that the battery intends to sell at hour $j$ (short position), $\mathsf{bs}_{ij}$ is the element of an indicator matrix which tracks the timing of spread trades (at row $i$, col $j$).

subject to:

$$\mathsf{b}_i+\mathsf{s}_i \le 1 \quad \forall i \in N \quad (2)$$
$$\mathsf{bs}_{ij} = \mathsf{b}_i+\mathsf{s}_j \quad \forall i \in N,\ \forall j \in N \quad (3)$$
$$\mathsf{bs}_{ij} \ge 0 \quad \forall i \in N,\ \forall j \in N \quad (4)$$
$$n = \sum_i \mathsf{b}_i \quad (5)$$
$$n = \sum_i \mathsf{s}_i \quad (6)$$

with $SoC$ constraints:

$$SoC_i = \begin{cases} SoC_s + \mathsf{b}_i - \mathsf{s}_i, & if\ i = 1 \\ SoC_{i-1} + \mathsf{b}_i - \mathsf{s}_i, & if\ i \neq 1 \end{cases} \quad (7)$$
$$SoC_{min} \le SoC_i \le SoC_{max} \quad \forall i \in N \quad (8)$$

where $SoC_i$ is the State of Charge of battery at hour $i$, $SoC_s$ is the State of charge of the battery at the starting time point.

and VaR constraints:

$$\eta|\hat{q}_{xx}^{(i,j)}| - c > 0 \quad \forall i = 0 : N-1, j = 0 : N-1 \quad (9)$$

where $q_{xx}^{(i,j)}$ is the xxth quantile of the spread density for $Y^{(i,j)}$ (e.g. 05th or 95th).

The objective function given in Equation (1) maximizes the daily profit contribution from the spread trades that involves buying at hour $i$ and selling at hour $j$. The trade decisions are based on the expected spread values from the $24 \times 24$ spread matrix. Equation (2) refers to the constraint where buying and selling cannot happen at the same time instant $i$. Equation (3) creates an indicator matrix to keep track of the combination of hours $i$ and $j$ where a spread trade takes place. The elements of the indicator matrix must be greater or equal to zero as per Equation (4). Equations (5) and (6) correspond to the number of buy and sell trades possible in a day, respectively. The $SoC$ related constraints of the battery are summarized in Equations (7) and (8). Equation (9) is the VaR constraint to make sure that the buy and sell trades are at least as profitable as the transaction costs, respectively.

## 5. Data and Density Forecasts for German Application

All spread pairs of hourly prices from the day-ahead auction (a total of 276) are forecast directly and separately as density functions. This is motivated by the requirement to take into account the trader's risk aversion. If the battery operator were not risk averse, the optimization could be based simply upon expected values and hourly forecasts would be used. In the current formulation, the $q_{05}$ and $q_{95}$ value-at-risk quantiles for each spread are required. Since the hourly prices are not independent and not normal, these densities will generally be intractable as the difference of pairs of hourly price densities. Hence they are estimated directly. The same dataset (German day-ahead prices from 2012–2017) and modeling (multifactor skew-t) are used as in Abramova et al. [22]. Since that work developed and validated the econometrics in detail, those estimated densities can be used as state-of-the-art.

Figure 2 shows three spread densities from 2017, for the example hours 00–08, 08–12 and 16–20, which are not Gaussian (note that spreads between other price hours, which are not shown, also do not have a normal distribution). The spreads are computed as earlier hour price minus later hour price. It is clear that a flexible density function is required to capture the varying shapes and the tails for $q_{05}$ and $q_{95}$ quantiles in particular. Therefore, modeling skewness is crucial and Figure 3 shows the average spread skewness from German data for 2012–2017, and how positive and negative skewness occur at different times of the day. The skewness depends upon wind and solar output and therefore upon weather forecasts. With high wind and/or solar output, the excess generation causes prices to become very low, hence the negative skewness. With high demand and low renewable outputs, there is scarcity and high prices, hence the positive skewness.

The hourly day-ahead prices, wind forecasts, solar forecasts and actual load data were downloaded from https://data.open-power-system-data.org (accessed on 25 October 2017) for the period of 1 January 2012–31 March 2017. Other variables considered in the modeling were the ARA 1-month-forward benchmark index for steam coal (single price per month) and the GPL natural gas day-ahead forward (single price per day) for gas. The weekly seasonality and holidays were accounted for as a single dummy variable, taking on value of 1 for Saturday/Sunday and the main German state holidays.

The available spread data are denoted with $\boldsymbol{Y} \in \mathbb{R}^{1917\times 276}$ out of which 383 observations were held out for out-of-sample forecasting. The day-ahead spreads, $Y_t^{(s)}$, were predicted by density functions, the four parameters of which depended dynamically upon: (1) spread of the lagged intraday electricity price, (2) Gaspool forward daily gas price, (3) ARA forward daily coal price, (4) spread of wind day-ahead forecast, (5) spread of solar day-ahead forecast, (6) dummy variable given value of 1 for holidays/weekends, (7) spread of the day-ahead total load forecast, and (8) an interaction load variable, calculated as $Load_{spread} * \frac{1}{2}(Load_{earlierHr} + Load_{laterHr}) = \frac{1}{2} * (Load^2_{earlierHr} - Load^2_{laterHr})$ i.e., average of the load for the two hours from which the spread is calculated, multiplied by the load spread obtained for those hours.

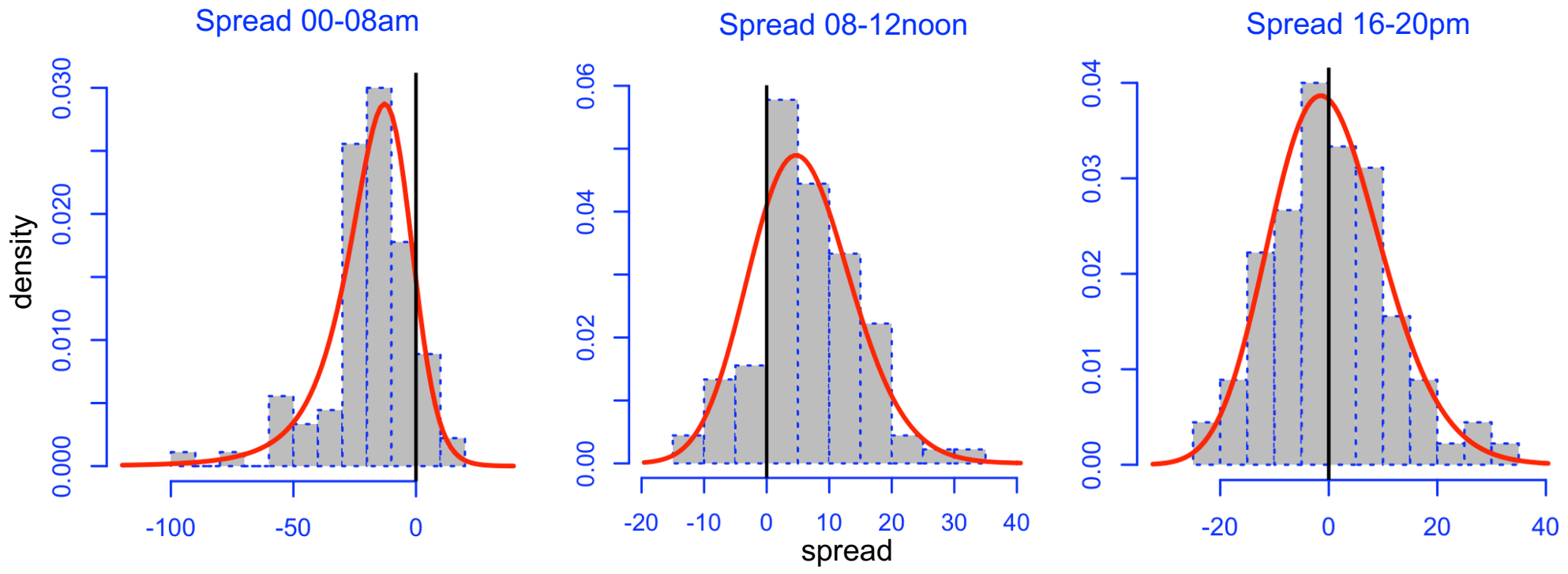

**Figure 2.** German spread densities in 2017.

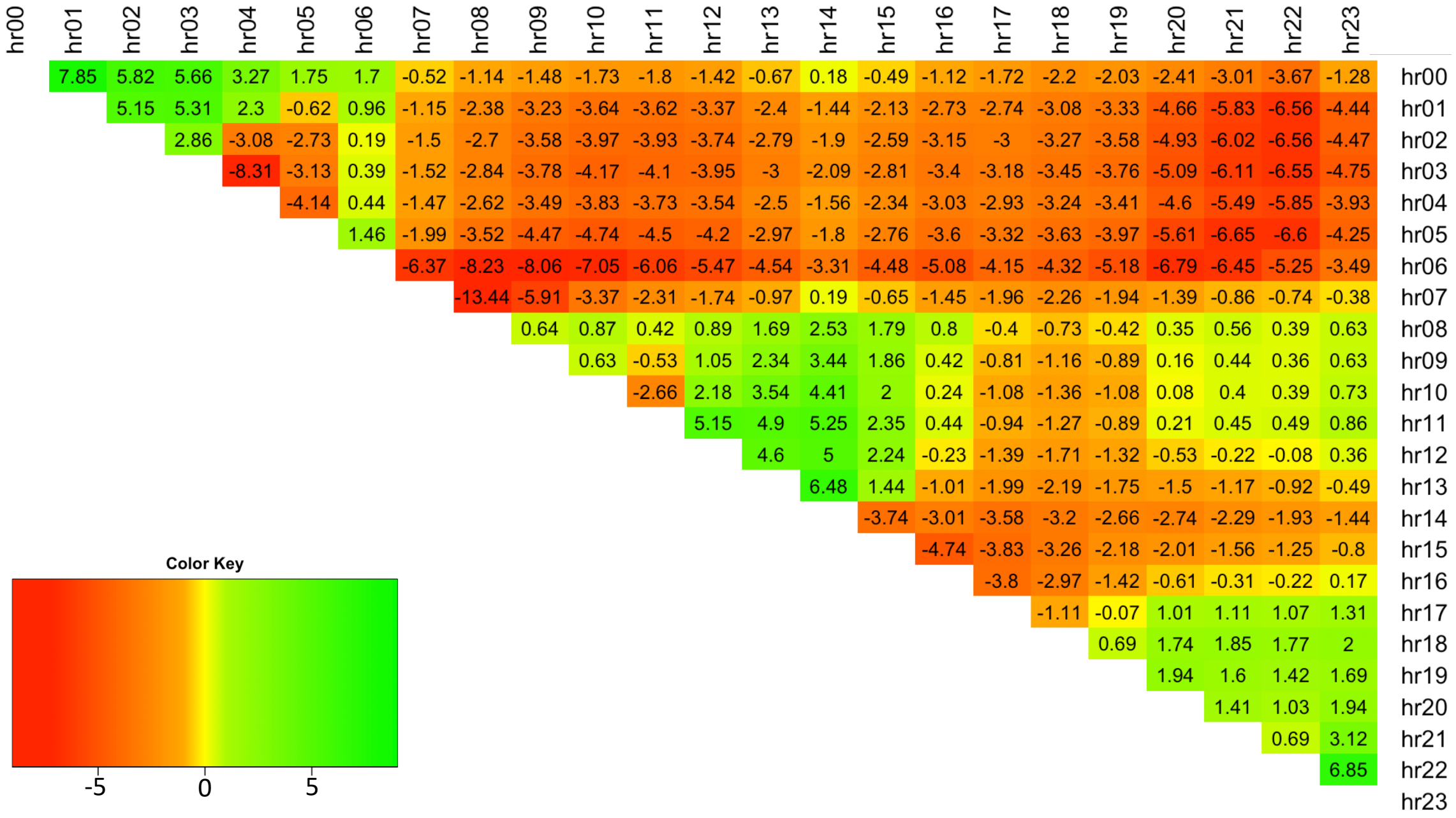

**Figure 3.** Average skewness per spread.

The above model for forecasting all the day-ahead spreads was specified and estimated on four years of data. From a range of functional forms that fit the data well, six four-parameter density functions were found to be appropriate (see Abramova et al. [22] for more details). In general, the skew-t was preferred, but not always. At each point in time adaptive re-specification and re-estimation was done to find which density function was performing the best for each spread. The results of the estimation were the four parameters $(\mu, \sigma, \nu, \tau)$ of every specific spread density specified as a dynamic function of explanatory variables. This resulted in an upper triangular matrix of density estimates for each spread, where each matrix element contained information on the six distributions used and the four estimated parameters describing each distribution for the data relating to that spread only. The distribution which was found to fit each spread data best, was chosen as the 'best distribution' (see Abramova et al. [22] for further details). The density functions have analytical solutions that allowed the $q_{05}$ and $q_{95}$ to be determined from their functional forms and the four forecast parameters. Thus, the modeling facilitated conditional estimates of the spread expectation and the tail risks based upon the exogenous day-ahead forecasts for wind, solar and other variables through the multifactor representation of each of the four parameters of the spread densities. The out-of-sample forecasts were then created for a further year and a half (383 days) to facilitate a backtesting simulation of trading. The optimization model of the previous section was then applied to the backtesting sample to determine the optimal operational schedule ahead of the auction. When the auction prices were subsequently revealed, the actual net payoff for the daily trading was determined.

Regarding the parameters, a number of assumptions have been made and sensitivity scenarios were applied. Consistently with other researchers, a 92% roundtrip efficiency, $\eta$, has been used for the battery storage cycles in this paper, i.e., 8% of the energy is lost in each cycle [11]. Recall that $c$ is the fixed roundtrip transaction cost for a storage facility in charging and discharging, which is distinct from the power loss in the charging discharging cycle. This covers the balancing, distribution, transmission, tax, trading, levies and other use of system costs for a facility wishing to operate in the wholesale market. The use of system and other non-technical transaction costs vary considerably according to national regulations and how the battery is connected to the system. If it is large and connected to the high-voltage transmission grid, in GB it is treated as a generator (despite also consuming) and will pay generators' transmission connection charges (contrast Germany where batteries are treated as consumption). However, if it is small and connected behind a retailer's meter, it will be subject to all the retail supply costs including distribution network charges and consumption levies. Due to the complexity of these costs, operators tend to have a fixed hurdle value for the roundtrip transaction cost, $c$, even though some of it is proportional to the wholesale selling price [4]. Estimates of these costs vary widely in practice and so for a methodological comparison $c = 5$ and 10 Euro/MWh were considered. This research investigates the optimal starting and finishing (by Property 4) SOC of the battery by backtesting for the two logical values, $b = \{0, 1\}$, and selected the one with the highest expected profit. The starting level of $b = 0$ was found to be optimal. A nominal 1 MWh fast charging battery is used, which is assumed to be fully (dis)chargeable within 1 h. This hourly dynamic constraint turned out to be less critical than might have been thought, as the charging and discharging events were generally well spaced out in the day, but it does make the calculation slightly easier if these events do not span two or more pricing periods.

## 6. Backtested Results

Thus, based on the forecast spread densities, the optimal trades were identified and then backtested against the actual spread values realized in the market. Ahead of the auction the expected net payoff, $E(\pi_t)$, on day $t$ was estimated with the expected spread, $E(Y_t^{(i,j)})$, as:

$$E(\pi_t) = \sum_{i=0}^{N-1} \sum_{j=0}^{N-1} \left[ \frac{\eta}{2} E(Y_t^{(i,j)})(\mathsf{bs}^*_{ij,t} - 1)(\mathsf{b}^*_{i,t} + \mathsf{s}^*_{j,t}) \right] - n * c \tag{10}$$

where $\mathsf{bs}^*_{ij,t}$, $\mathsf{b}^*_{i,t}$ and $\mathsf{s}^*_{j,t}$ are the optimal results from Equation (1) for a given day $t$.

Repeating over all the 383 days, held back for out-of-sample testing, provides a set of expected net payoffs which can be compared with actual net payoffs, $\pi_t$, based on the actual spreads, $y_t^{(i,j)}$, obtained for those trades:

$$\pi_t = \sum_{i=0}^{N-1} \sum_{j=0}^{N-1} \left[ \frac{\eta}{2} y_t^{(i,j)}(\mathsf{bs}^*_{ij,t} - 1)(\mathsf{b}^*_{i,t} + \mathsf{s}^*_{j,t}) \right] - n * c \tag{11}$$

For computational simplicity, robustness and transparency, optimization was performed by means of an exhaustive search heuristic, as described in the Appendix A. With only 276 spreads to search for one day, the exhaustive search heuristic on the spreads avoids the need to estimate a price level model to implement the turning points heuristic of Section 3.

### 6.1. One Trade Per Day

Initially a single daily cycle mode of operation is considered and the optimal trade to be performed on each day is found using an exhaustive state space search over all available trade scenarios, performed using Algorithm A1. Thus, at each time step, $t = 1, \ldots, 383$, i.e., each one day of the backtesting period, $s = 1, \ldots, 276$ potential trades based on the forecasted spread densities are considered, where for each trade the possibility of making a profit of at least $c$ Euro/MWh is examined with 95% confidence. Please note that trading could be with either a sufficiently large positive spread (sell first then buy) or a negative spread (buy then sell). For those trades which pass this value-at-risk constraint, the one with maximum expected profit is selected.

Thus, if $E(Y_t^{(s)}) < 0$ the 95th quantile is accessed so that body of the distribution is to the left-hand side of the critical value (i.e., later price is higher, hence charge then discharge). The expected payoff is the difference between the expected value (accounted for battery inefficiency) and the total roundtrip cost, multiplied by the level of battery charge available to trade (e.g., $b$ if the trade is to sell first). This is referred to as the expected payoff for trade $s$ on day $t$, $E(\pi_t^{(s)})$,

$$E(\pi_t^{(s)}) = \begin{cases} \left(\eta E(Y_t^{(s)}) - c\right)b & \text{if } \eta E(\hat{Y}_t^{(s)}) > 0 \\ \left(\eta \left|E(Y_t^{(s)})\right| - c\right)(1-b) & \text{if } \eta E(\hat{Y}_t^{(s)}) < 0 \end{cases} \tag{12}$$

Repeated over all possible trades $s = 1, \ldots, 276$ for each day, and all out-of-sample days, $t$, this provides the matrix of expected payoffs in the out-of-sample data, $t = 1, \ldots, 383$, which is referred to as the expected payoffs matrix, $E(\mathbf{\Pi}) \in \mathbb{R}^{276 \times 383}$. To backtest the data, the chosen trades are compared with actual outcomes to create a realized strategy payoff vector, $\boldsymbol{\pi} \in \mathbb{R}^{383}$, the elements of which ($\pi_t$) represent the realized strategy daily payoffs, $t = 1, \ldots, 383$.

The following quantities are reported for each starting battery level, $b \in \{0, 1\}$: total realized strategy payoff over the backtest period $\pi_{total}$ (Equation (13)), average realized strategy payoff over the backtest period $\overline{\pi}$ (Equation (14)), standard error of the realized strategy payoffs $s^{\pi}$ (in which $sd_{\pi}$ is the sample standard deviation obtained from the payoff vector, $\boldsymbol{\pi} \in \mathbb{R}^{383}$) (Equation (15)), number of trades which resulted in a loss after all costs are taken into account $n_l$, (Equation (16)), total monetary value resulting from loss days $l$ (Equation (17)), and average loss $\bar{l}$ (Equation (18)).

$$\pi_{total} = \sum_{t=1}^{383} \pi_t \tag{13}$$

$$\overline{\pi} = \frac{1}{383} \sum_{t=1}^{383} \pi_t \tag{14}$$

$$s^{\pi} = \frac{1}{\sqrt{383}} sd_{\pi} \tag{15}$$

$$n_l = \sum_{t=1}^{383} 1 \qquad \text{if } \pi_t < 0 \tag{16}$$

$$l = \sum_{t=1}^{383} \pi_t \qquad \text{if } \pi_t < 0 \tag{17}$$

$$\bar{l} = \frac{l}{n_l} \tag{18}$$

Algorithm A1 summarizes the pseudocode. Please note that the predictive model for each spread at each time period, $\widehat{M}_t^{(s)}$, is adaptively chosen from several candidate models, while the $\widehat{\boldsymbol{\theta}}_t^{(s)} = [\hat{\mu}_t^{(s)}, \hat{\sigma}_t^{(s)}, \hat{\nu}_t^{(s)}, \hat{\tau}_t^{(s)}]^T$ are the predicted parameters of the best distribution, as described in Abramova et al. [22].

### 6.2. Two Trades Per Day

Undertaking an exhaustive search for two trades per day builds upon this approach. The optimum two trade per day schedule is established for each initial battery level $b = \{0, 1\}$ and minimum roundtrip cost $c = 5, 10$ Euro/MWh, analogously to the single trade search. As before, the forecasted payoff values are calculated, $E\left(\pi_t^{(s)}\right)$, which correspond to performing a single full trade for spread number $s = 1, \ldots, 276$ at test time step $t$. Next the overall expected strategy payoff for carrying out two trades, $E(\mathbf{\Pi}^{tot}) \in \mathbb{R}^{276\times383}$, is calculated, where the second trade is carried out *only* in the case when a feasible (profitable, after roundtrip transaction costs and battery efficiency are taken into account) second trade spread exists. Then up to two trades which yield the highest overall expected payoff at each test time step are selected. In the case where algorithm forecasts the second trade to not be profitable, it would default to trading only a single cycle in that day. The overall algorithm is reported in two stages in Algorithms A3 and A4. The summary statistics of the results are reported analogously to the one trade case. In principle, this sequential search approach could be extended to three trades per day however it was not found to be optimal for the given data.

## 7. Results

The trading results for the smaller sensitivity roundtrip cost of $c = 5$ Euro/MWh are considered first, the results of which are reported in Table 1 using summary statistics given in Equations (13)–(18) and obtained for estimated models using the best distributions (termed BD) for each forecast (out of the $276 * 383 = 105{,}708$ spread forecasts over the investment horizon, there were 1795 and 4159 spreads for which the means of BD and ND, respectively, were not forecast successfully) and the benchmark normal distribution (termed ND). On any day where a profitable trade was not forecasted, a position was not put on (from possible 383 backtested days, a profitable trade was not forecast on the following number of days under BD models: c = 5, b = 0 *8* days; c = 5, b = 1 *107* days; c = 10, b = 0 *146* days; c = 10, b = 1 *291* days; while under ND models: c = 5, b = 0 *7* days; c = 5, b = 1 *104* days; c = 10, b = 0 *187* days; c = 10, b = 1 *325* days).

**Table 1.** Single trade scenario, cost $c = 5$ Euro/MWh, battery efficiency $\eta = 0.8$. Results for models estimated under best distribution vs. the benchmark normal.

| **Best Dist.** | | | **Normal Dist.** | | |
|---|---|---|---|---|---|
| $b$ | **0** | **1** | $b$ | **0** | **1** |
| $\pi_{total}$ | 4853.8 | 3186.4 | $\pi_{total}$ | 4828.8 | 2942.4 |
| $\overline{\pi}$ | 12.94 | 11.54 | $\overline{\pi}$ | 12.84 | 10.55 |
| $s^{\pi}$ | 0.61 | 0.72 | $s^{\pi}$ | 0.65 | 0.79 |
| $n_l$ | 3 | 8 | $n_l$ | 15 | 31 |
| $l$ | −2.2 | −6.4 | $l$ | −33.4 | −66.2 |
| $\bar{l}$ | −0.73 | −0.80 | $\bar{l}$ | −2.23 | −2.14 |

The total realized strategy payoffs across the two battery levels for the normal and non-normal models is very similar and on average the best initial charge setting was zero. However, from a risk perspective, at the zero setting, the ratio of total number of loss days is 5 times more for the normal, with the monetary value of these losses 15 times higher over models based on the best chosen distributions. Furthermore, when the higher value of 10 Euro/MWh roundtrip cost is considered the normal distribution is substantially outperformed on total realized strategy payoff as well. The total realized strategy payoff results for 10 Euro/MWh roundtrip cost trading scenario are reported in Table 2.

**Table 2.** Single trade scenario, cost $c = 10$ Euro/MWh, battery efficiency $\eta = 0.8$. Results for models estimated under best distribution vs. the benchmark normal.

| Best Dist. | | | Normal Dist. | | |
|---|---|---|---|---|---|
| **b** | **0** | **1** | **b** | **0** | **1** |
| $\pi_{total}$ | 2702 | 1318.4 | $\pi_{total}$ | 2037.6 | 664.8 |
| $\overline{\pi}$ | 11.40 | 14.33 | $\overline{\pi}$ | 10.40 | 11.46 |
| $s^{\pi}$ | 0.825 | 1.66 | $s^{\pi}$ | 1.04 | 1.60 |
| $n_l$ | 15 | 2 | $n_l$ | 25 | 2 |
| $l$ | −14 | −3.2 | $l$ | −82 | −5.6 |
| $\bar{l}$ | −0.93 | −1.6 | $\bar{l}$ | −3.28 | −2.8 |

These backtested simulation results reveal the advantage of using skew-type and similar distribution-based models: the average total realized strategy payoff is significantly better than normal at 95% confidence, and the best initial battery charge on average is at the 0% charge level (i.e., fully discharged). The total realised strategy payoff over the tested period for 10 Euro/MWh cost and battery level $b = 0$ is 25% higher using non-normal densities over the normal. The ratio of total number of loss days for normal to skew models is 1.6, with the monetary value of losses being 5.9 times higher for the normal types.

Now the results from an algorithm which seeks to operate with up to two trades per day are discussed. The roundtrip cost of 5 Euro/MWh is considered first, the results of which are reported in Table 3. The optimal battery starting/finishing state of 0 (fully discharged) is examined. The results show that the skew type has selected more two-trade days throughout the test horizon as compared to the normal distribution, therefore yielding a substantially better performance. A much higher profit compared to the single trade scenario can be seen (where the average payoff of skew densities is now 7% larger, with the standard error almost unchanged and the total loss unchanged, therefore the selected second trades on average increased the payoff). Further analysis of the 87 two battery operation schedules shows that generally the two-trade option results in a higher profit than the one trade option.

**Table 3.** Two-trade scenario, cost $c = 5$ Euro/MWh, battery efficiency $\eta = 0.8$. Results for models estimated under best distribution vs. the benchmark normal.

| Best Dist. | | | Normal Dist. | | |
|---|---|---|---|---|---|
| **b** | **0** | **1** | **b** | **0** | **1** |
| $\pi_{total}$ | 5200.8 | 3357.8 | $\pi_{total}$ | 5039.4 | 3089.2 |
| $\overline{\pi}$ | 13.87 | 12.17 | $\overline{\pi}$ | 13.4 | 11.1 |
| $s^{\pi}$ | 0.63 | 0.74 | $s^{\pi}$ | 0.67 | 0.79 |
| $n_l$ | 3 | 8 | $n_l$ | 14 | 31 |
| $l$ | −2.2 | −6.4 | $l$ | −33.2 | −66.2 |
| $\bar{l}$ | −0.73 | −0.80 | $\bar{l}$ | −2.37 | −2.13 |
| $n_1$ | 288 | 245 | $n_1$ | 321 | 249 |
| $n_2$ | 87 | 31 | $n_2$ | 55 | 30 |

The most profitable trade was at test time step $t = 56$, where only a single trade is selected for the spread between hours 14–21, yielding a profit of €115. On this day, with starting battery level of 0, the only feasible trade is for a charge at 14.00 h and discharge at high point of 21.00 h (between those hours the electricity price profile is a monotonically increasing function). Figure 4 shows this trade, which was selected both by single and double trade algorithms (note that in this case the double trade algorithm defaulted to trading only once, since no further trade was deemed profitable).

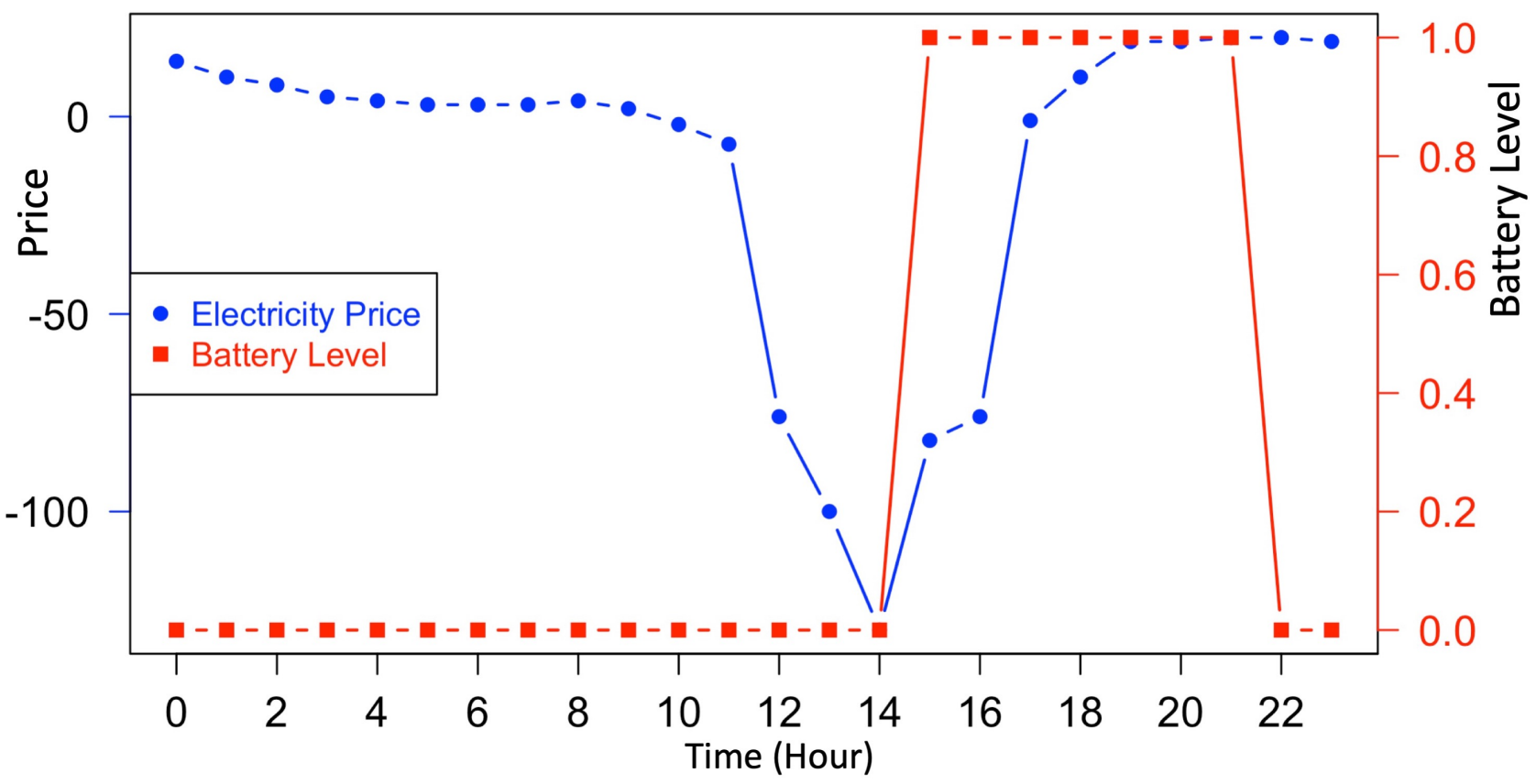

**Figure 4.** Price and battery profiles for test time step $t = 56, b = 0, c = 5, \eta = 0.8$. Single trade selected by both algorithms, the spread between hours 14–21.

The second most profitable trade was at test time step $t = 317$, where different trades were selected by the single and double trade algorithms. Figure 5 depicts the price and battery level profiles, where for a two-trade scenario Figure 5a shows the total profit obtained was €82 (trade 1 selecting spread 4–9: €73.4 + trade 2 selecting spread 12–18: €8.6), while the corresponding single trade, reported for comparison in Figure 5b, used the spread 00–18 yielding €60.6. Hence demonstrating a clear advantage of allowing two cycles of trading to be performed intraday.

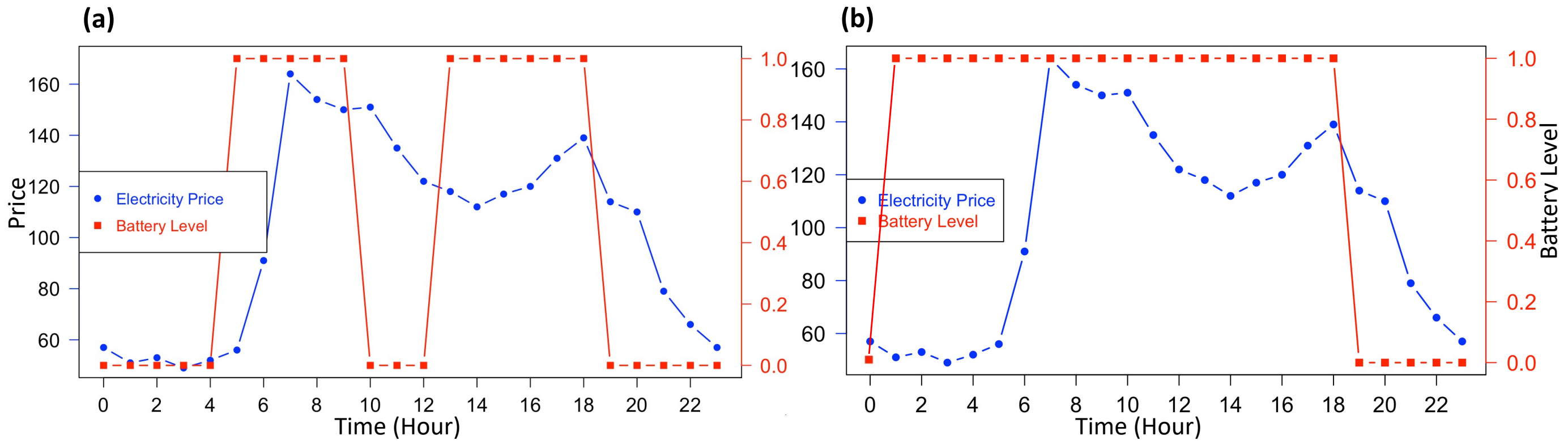

**Figure 5.** Price and battery profiles for test time step $t = 317, b = 0, c = 5, \eta = 0.8$ (**a**) two trades using spreads 4–9, 12–18; (**b**) single trade using spread 00–18.

The two trades with roundtrip cost of 10 Euro/MWh are considered next. The results are reported in Table 4. The cost in this case is too high for a second profitable trade, and the results show that the skew type traded twice only one time at $b = 0$ and normal distribution only twice, while not improving on the total payoff as compared to single trade scenario. At $b = 1$ both the strategy and benchmark only traded once per day, showing results unchanged from the single trade scenario discussed above.

The key results of Tables 1–4 are summarized in Figure 6 where each bar shows results for skew and normal type overlaid onto a single bar. The bars are obtained as follows: in Figure 6a(i) the single trade algorithm under cost $c = 5$ has bar length reflecting the total loss of −€33.4 obtained for the normal distribution (brown color). The total loss obtained with the skew-type algorithm is overlaid onto the normal bar plot to show the value of −€2.2 (green color), as seen in Table 1.

**Table 4.** Two-trade scenario, cost $c = 10$ Euro/MWh, battery efficiency $\eta = 0.8$. Results for models estimated under best distribution vs. the benchmark normal.

| Best Dist. | | | Normal Dist. | | |
|---|---|---|---|---|---|
| $b$ | 0 | 1 | $b$ | 0 | 1 |
| $\pi_{total}$ | 2696 | 1318.4 | $\pi_{total}$ | 2039.2 | 664.8 |
| $\overline{\pi}$ | 11.38 | 14.33 | $\overline{\pi}$ | 10.40 | 11.46 |
| $s^{\pi}$ | 0.825 | 1.66 | $s^{\pi}$ | 1.06 | 1.60 |
| $n_l$ | 15 | 2 | $n_l$ | 25 | 2 |
| $l$ | −14 | −3.2 | $l$ | −82 | −5.6 |
| $\bar{l}$ | −0.93 | −1.6 | $\bar{l}$ | −3.28 | −2.8 |
| $n_1$ | 236 | 92 | $n_1$ | 194 | 58 |
| $n_2$ | 1 | 0 | $n_2$ | 2 | 0 |

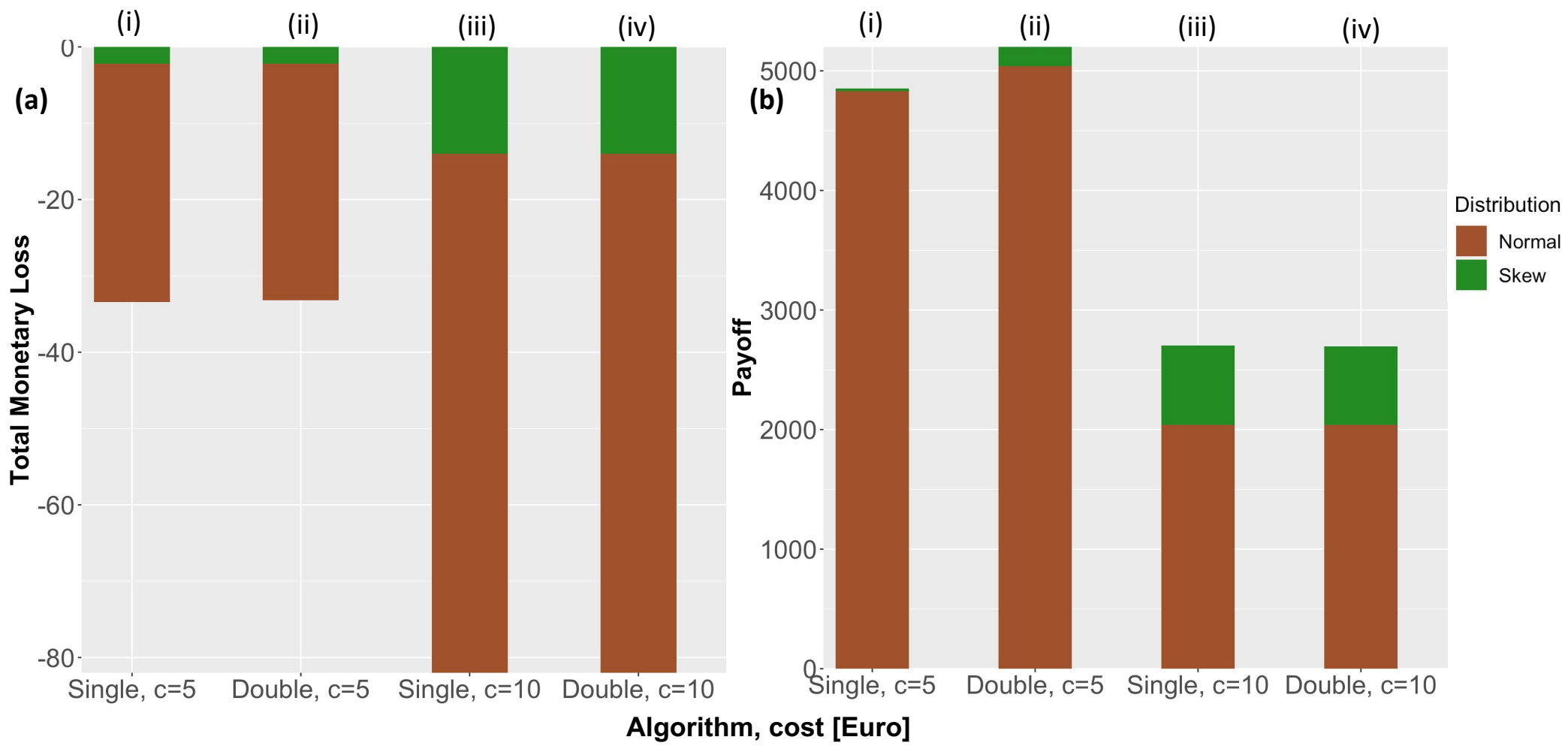

**Figure 6.** Key results of single trade and double trade algorithms under initial battery level $b = 0$, transaction costs, $c = \{5, 10\}$ and normal/skew-type distributions; (**a**) total monetary loss across the backtest horizon, (**b**) total payoff accumulated over the backtest horizon.

Figure 6a depicts the total monetary loss for the algorithms under normal and skew-type distributions and Figure 6b shows the total payoff, both obtained over the backtesting horizon. Examining the transaction costs of $c = 5$ (bars $i - ii$ in each subplot), it is observed that for the single trade algorithm (Figure 6b(i)), the payoff of the normal and skew-type algorithms are similar while the total monetary loss (Figure 6a(i)) of the normal is very large as compared to the skew-type. The double trade algorithm increases the profits slightly (Figure 6b(ii)) while keeping the same profile for the total monetary loss (Figure 6a(ii)), showing a clear advantage of the skew-type distributions. As the transaction costs are increased to $c = 10$ the payoff decreases for both normal and skew type (both for single and double algorithms (Figure 6b(iii,iv)) since less trading happens, while the total monetary loss increases (Figure 6a(iii,iv)). However, the clear superiority of the skew-type is demonstrated through a much smaller total loss (Figure 6a(iii,iv)) compared to normal.

## 8. Conclusions

Within liberalized power markets, practical evidence points to commercial battery operators seeking revenues through various trading opportunities and network services. Within these portfolios, one of the main revenue streams is derived by seeking to arbitrage the hourly price spreads in the day-ahead auction. The optimal daily approach to this is challenging if risk is a consideration to the operations manager and trader. If risk management compliance requires a VaR constraint, then a probabilistic formulation of the arbitrage spreads becomes necessary. This requires density functions and because the hourly prices are not independent, this precludes creating spread densities from the difference of price densities as forecasted directly. This research is the first to formulate forecasts of all 276 intraday hourly spreads using flexible four-parameter distributions to reflect the way in which the skewness and the tails of the densities change during the day and over time. It produces dynamic conditional parameter estimates driven by exogenous factors, most prominently the day-ahead demand, wind and solar forecasts. These forecasts fitted well in backtesting and supported the optimal daily scheduling of a storage facility, operating on a single cycle and on two cycles. This paper finds that depending upon weather forecasts and the roundtrip transaction costs of a charging/discharging cycle as well as considering battery inefficiency costs, a two cycle operation is often more profitable than the conventional single cycle. This is likely to become more so as distributed solar generation increases during the middle of the day and therefore depresses prices to create deep midday troughs. The model outperformed benchmark comparisons to a normal density model. The optimal choice of spreads to trade varies daily and the need to use forecasts in well-specified models is apparent, as is the delicate balance between expected profits and risk. Furthermore, the computational nature of the modeling presented would lead naturally to the prospect for algorithmic trading by battery asset owners, which may be more practical for small enterprises than outsourcing their trading to larger service providers.

It is important to review the proposed model's limitations and further research. The rigorous approach of computing and selecting the best distribution for each spread from the possible choice of six distributions, results in a high computational demand. To mitigate this limitation, the parameter estimation phase can be simplified using a skew-t type 5 distribution for estimating the parameters for all spreads. Further investigation of the robustness and generalization of this assumption would be useful. Finally, the initial perspective of spread trading being just one of several operational monetization of a battery asset is re-visited. In practice the asset owners are likely to use several approaches. In particular, batteries can provide the valuable fast reserve services for balancing by the system operator and the flexibility services increasingly required by local distribution network operators. The daily horizon of pure, fully balanced spread trading which results from the need to open and close a trade within the same day-ahead auction, is efficient from a risk management perspective and forms an attractive ingredient to more flexible portfolio operations. However, as this paper has emphasized, it comes with the constraint of a closed day-ahead trading position. Whether more dynamic, multiperiod, continuous look-ahead trading, which does not use fully balanced spreads but a sequence of open positions, provides better expected returns at lower risk, will depend upon circumstances. These circumstantial features may include the higher bid-offer spreads in continuous trading, compared to day-ahead auctions, as well as the lower predictability and higher risks of longer-term positions than day ahead.

**Author Contributions:** Both authors contributed equally in all stages of the research. All authors have read and agreed to the published version of the manuscript.

**Funding:** This research was funded by the UK EPSRC research council as part of the AGILE project (EP/S003088/1) on digital aggregators.

**Data Availability Statement:** The German hourly day-ahead electricity price, actual total load, wind and solar forecast data were downloaded from the Open Power System Data website https://data.open-power-system-data.org (accessed on 25 October 2017) for the period of 1 January 2012 to 31 March 2017. The steam coal ARA 1-month-forward benchmark index was used for the steam coal data (one price per month) and the Germany Gaspool (GPL) natural gas day-ahead forward (one price per day)was used for gas.

**Acknowledgments:** The authors wish to acknowledge the help of Priyanka Shinde in developing the research review of stochastic optimization methods for batteries.

**Conflicts of Interest:** The authors declare no conflict of interest.

## Abbreviations

The following abbreviations are used in this manuscript:

| | |
|---|---|
| DA | Day-Ahead |
| PJM | Pennsylvania, Jersey, Maryland Power Pool |
| GAMLSS | Generalized Additive Model for Location, Scale and Shape |
| ARIMA | Autoregressive Integrated Moving Average |
| PV | Photovoltaic |
| SDP | Stochastic Dynamic Programming |
| ADP | Approximate Dynamic Programming |
| DSDP | Discretized SDP |
| MDP | Markov Decision Process |
| SoC | State of Charge |
| $\text{SoC}_s$ | Starting State of Charge |
| $\text{SoC}_f$ | Finishing State of Charge |
| P1 / P2 / P3 / P4 | Property number 1/2/3/4 |
| VaR | Value-at-Risk |
| ARA | Amsterdam-Rotterdam-Antwerp |
| GPL | Germany Gaspool |
| GB | Great Britain |
| MWh | Megawatt Hour |
| BD | Best Distribution (skew type) |
| ND | Normal Distribution |

## Nomenclature

The following nomenclature was used in this manuscript:

| | |
|---|---|
| $i$ | Hourly index, $0 \ldots 23$, corresponding to the buy trade |
| $j$ | Hourly index, $0 \ldots 23$, corresponding to the sell trade |
| $Y^{(i,j)}$ | Random variable (r.v.) representing individual spread price, element of a matrix $\in \mathbb{R}^{24\times 24}$ (element of row $i$, col $j$) |
| $s$ | Spread number (flattened $24 \times 24$ hourly matrix), element of a vector $\in \mathbb{R}^{276}$ |
| $Y_t^{(s)}$ | Shorthand notation for r.v. $Y^{(i,j)}$ i.e., spread price for spread $s$ at time step $t$ |
| $y_t^{(s)}$ | Realized spread price for spread $s$ at time step $t$ |
| $\boldsymbol{Y}$ | Full dataset of spread prices $\in \mathbb{R}^{1917\times 276}$, last 383 observations used for out-of-sample |
| $E(Y_t^{(s)})$ | Expected price of spread $s$ at time $t$ (forecasted / fitted spread price value) |
| $q_{05}^{(i,j)}$ | 5th quantile of the spread density for $Y^{(i,j)}$ |
| $q_{95}^{(i,j)}$ | 95th quantile of the spread density for $Y^{(i,j)}$ |
| $\eta$ | Battery efficiency for roundtrip spread trade |
| $c$ | Transaction costs for roundtrip spread trade |
| $b$ | Starting/finishing state of charge of the battery level, $\in \{0, 1\}$ |
| $\mathsf{b}_i$ | Amount of energy, $\geq 0$, to be bought by the battery at hr $i$ (long position) |
| $\mathsf{s}_j$ | Amount of energy, $\geq 0$, that the battery intends to sell at hr $j$ (short position) |
| $\mathsf{bs}_{ij}$ | Element of indicator matrix that tracks timing of spread trades (row $i$, col $j$) |
| $\mathsf{b}_i^*$ | Optimal value of $\mathsf{b}_i$ |
| $\mathsf{s}_j^*$ | Optimal value of $\mathsf{s}_j$ |
| $\mathsf{bs}_{ij}^*$ | Optimal value of $\mathsf{bs}_{ij}$ |
| $N$ | Number of hours in a day available for the battery to trade, $N = 24$ |
| $n$ | Number of spread trades per day |
| $n_1$ | Number of days with 1 spread trade per day, from 383 out-of-sample data |
| $n_2$ | Number of days with 2 spread trades per day, from 383 out-of-sample data |
| $n_l$ | Number of loss days after costs are taken into account, from 383 out-of-sample data |
| $\widehat{M}_t^{(s)}$ | Learnt model for spread $s$ at $t$ specifying the best distribution and its parameters |
| $\widehat{\boldsymbol{\theta}}_t^{(s)}$ | Estimated parameters of the best skew-type of distribution for spread $s$ at time step $t$ |
| $E(\pi_t^{(s)})$ | Expected net payoff for spread $s$ of day $t$, $\in\mathbb{R}$ |
| $E(\pi_t^{(s)})$ | Expected net payoff of day $t$, $\in\mathbb{R}$ |
| $\pi_t$ | Realized net payoff for spread $s$ of day $t$ (actual profit / loss), $\in \mathbb{R}$ |
| $\boldsymbol{\pi}$ | Realized net payoff vector, $\in\mathbb{R}^{383}$ |
| $E(\boldsymbol{\Pi})$ | Expected payoffs matrix single trade scenario, $\in\mathbb{R}^{276\times383}$ |
| $E(\boldsymbol{\Pi}_t)$ | Vector of expected payoffs from $E(\boldsymbol{\Pi})$ at time $t$, (can be reshaped into $\in\mathbb{R}^{24\times24}$) |
| $\boldsymbol{\Pi}$ | Actual matrix of all possible payoffs, single trade scenario, $\in\mathbb{R}^{276\times383}$ |
| $\boldsymbol{\Pi}_t$ | Vector of realized strategy payoffs, single trade scenario, $\in\mathbb{R}^{276}$ at $t$ |
| $\pi_t^{(s)}$ | Element of actual payoff vector, $\boldsymbol{\Pi}_t$, i.e., payoff at $t$ for spread trade $s$, $\in\mathbb{R}$ |
| $\pi_{total}$ | Total realized strategy payoff over the backtest period, $\in\mathbb{R}$ |
| $\overline{\pi}$ | Average realized strategy payoff, of 383 out-of-sample data, $\in\mathbb{R}$ |
| $s^{\pi}$ | Standard error of the realized strategy payoffs, of 383 out-of-sample data, $\in\mathbb{R}$ |
| $sd_{\pi}$ | Sample std. dev. obtained from payoff vector, $\boldsymbol{\pi}$, of 383 out-of-sample data, $\in\mathbb{R}$ |
| $l$ | Total monetary value resulting from loss days, from 383 out-of-sample data, $\in\mathbb{R}$ |
| $\bar{l}$ | Average loss value resulting from loss days, from 383 out-of-sample data, $\in\mathbb{R}$ |
| $E(\boldsymbol{\Pi}^{tot})$ | Expected total payoff matrix, 2 trade scenario, $\in\mathbb{R}^{276\times383}$ |
| $E(\boldsymbol{\Pi}_t^{tot})$ | Expected total payoff vector at time step $t$, 2 trade scenario, $\in\mathbb{R}^{276}$ |
| $\boldsymbol{\Pi}^{tot}$ | Actual total payoff matrix, 2 trade scenario, $\in\mathbb{R}^{276\times383}$ |
| $E\left(\pi_t^{tot,(s)}\right)$ | Element of $E(\boldsymbol{\Pi}^{tot})$ for spread $s$ at time $t$, $\in\mathbb{R}$ |
| $\pi_t^{tot,(s)}$ | Element of $\boldsymbol{\Pi}^{tot}$ for spread $s$ at time $t$, $\in\mathbb{R}$ |
| $\boldsymbol{\pi^{tot}}$ | Realized total payoff vector, $\in \mathbb{R}^{383}$ |

## Appendix A. Algorithms

Pseudocode is written assuming R programming language (indexing starts from 1).

**Algorithm A1** One Trade Forecasted Strategy Payoff (Battery Level $b$ and Risk Criterion $c$)

Init strategy *expected* payoff matrix, $E(\boldsymbol{\Pi}) \leftarrow \mathbf{0} \in \mathbb{R}^{276\times383}$
**for** each time step (i.e., test point) $t = 1, \ldots, 383$ **do**
  **for** each hour spread $s = 1, \ldots, 276$ **do**
    Use model $\widehat{M}_t^{(s)}$ to extract forecasted parameters $\widehat{\boldsymbol{\theta}}_t^{(s)} = [\hat{\mu}_t^{(s)}, \hat{\sigma}_t^{(s)}, \hat{\nu}_t^{(s)}, \hat{\tau}_t^{(s)}]^T$
    **if** distribution parameters $\widehat{\boldsymbol{\theta}}_t^{(s)}$ were successfully forecasted **then**
      Extract calculated expected value of the spread, $E(Y_t^{(s)})$
      **if** $E(Y_t^{(s)}) > 0$ **then**
        Extract 5th quantile (i.e., 95% of dist. is on the right), $\hat{q}_{05,t}^{(s)}$, using $\widehat{\boldsymbol{\theta}}_t^{(s)}$
      **else**
        Extract 95th quantile (i.e., 95% of dist. is on the left), $\hat{q}_{95,t}^{(s)}$, using $\widehat{\boldsymbol{\theta}}_t^{(s)}$
      **if** $\eta|\hat{q}_{xx,t}^{(s)}| > c$ **then**
        **if** $E(Y_t^{(s)}) > 0$ **then**
          $E(\pi_t^{(s)}) \leftarrow \left(\eta E(Y_t^{(s)}) - c\right)b$, *expected* payoff for spread $s$ at $t$
        **else**
          $E(\pi_t^{(s)}) \leftarrow \left(\eta |E(Y_t^{(s)})| - c\right)(1-b)$, *expected* payoff for spread $s$ at $t$

**Algorithm A2** One Trade Actual Strategy Payoff (Battery Level $b$ and Risk Criterion $c$)

Init strategy *actual* payoff matrix $\mathbf{\Pi} \leftarrow \mathbf{0} \in \mathbb{R}^{276\times 383}$ payoff for each spread $s$ at each $t$
Init strategy *realized* payoff vector, $\boldsymbol{\pi} \leftarrow \mathbf{0} \in \mathbb{R}^{383}$
**for** each time step (i.e., test point) $t = 1, \ldots, 383$ **do**
  **for** each hour spread $s = 1, \ldots, 276$ **do**
    Extract actual (observed) spread price $y_t^{(s)}$
    **if** $y_t^{(s)} > 0$ **then**
      $\pi_t^{(s)} \leftarrow (\eta y_t^{(s)} - c)b$, element of $\mathbf{\Pi}_t$
    **else**
      $\pi_t^{(s)} \leftarrow \big(\eta |y_t^{(s)}| - c\big)(1-b)$, element of $\mathbf{\Pi}_t$
  $maxVal \leftarrow max_s E(\mathbf{\Pi}_t)$, extract max value from vector of *expected* payoffs at $t$
  **if** $maxVal \neq 0$ **then**
    $s' \leftarrow argmax_s \mathbf{\Pi}_t$, extract spread hour associated with max *expected* payoff at $t$
    $\pi_t \leftarrow \pi_t^{(s')}$ extract payoff from $\mathbf{\Pi}_t$ at $s'$, $t$; assigned as elem of *realised* payoff $\boldsymbol{\pi}$

**Algorithm A3** Two-Trade Forecasted Strategy Payoff (Battery Level $b$ and Risk Criterion $c$)

Init strategy *expected* total payoff matrix, $E(\mathbf{\Pi}^{tot}) \in \mathbb{R}^{276\times 383}$
Init strategy *actual* total payoff matrix, $\mathbf{\Pi}^{tot} \in \mathbb{R}^{276\times 383}$
**for** each time step (i.e., test point) $t = 1, \ldots, 383$ **do**
  **for** each spread hour $s = 1, \ldots, 276$ **do**
    **if** $E(\pi_t^{(s)}) > 0$, if exp. payoff for 1st spread $s$ at $t$ is profitable **then**
      Init $\boldsymbol{S} \leftarrow E(\mathbf{\Pi}_t) \in \mathbb{R}^{24\times 24}$ expected payoffs for 2nd trade; copy of 1st trade
      Set $\boldsymbol{S}$[ii-1:jj, :] $\leftarrow \mathbf{0}$; and $\boldsymbol{S}$[1:ii-1, ii:24] $\leftarrow \mathbf{0}$, only allow feasible 2nd trades
      `found` $\leftarrow$ `False`
      **while** not `found` **do**
        $maxVal \leftarrow \max_s \boldsymbol{S} \in \mathbb{R}_0^+$, max payoff value over all possible 2nd trades
        **if** $maxVal = 0$ **then**
          $E\left(\pi_t^{tot,(s)}\right) \leftarrow E(\pi_t^{(s)}) + 0$, *expected* total payoff of 1 trade only
          $\pi_t^{tot,(s)} \leftarrow \pi_t^{(s)} + 0$, *actual* total payoff comprised of 1 trade only
          trade `found` $\leftarrow$ `True`
        **else** Calculate total two trade payoff
          $s' \leftarrow \text{argmax}_s\, \boldsymbol{S}$ spread hour $i' - j'$ corresp. to max expected payoff
          $E\left(\pi_t^{tot,(s)}\right) \leftarrow E(\pi_t^{(s)}) + \boldsymbol{S}^{(s')}$
          $\pi_t^{tot,(s)} \leftarrow \pi_t^{(s)} + \pi_t^{(s')}$
          trade `found` $\leftarrow$ `True`

**Algorithm A4** Two-Trade Realized Strategy Payoff (Battery Level $b$ and Risk Criterion $c$)

Init strategy *realised* total payoff vector, $\boldsymbol{\pi}^{tot} \leftarrow \mathbf{0} \in \mathbb{R}^{383}$
Init $n \leftarrow 0$ number of loss days
Init $L \leftarrow [\,]$ array to store monetary value of each loss day
**for** each test point $t = 1, \ldots, 383$ **do**
  $maxVal \leftarrow \max_s E(\mathbf{\Pi}_t^{tot})$, select largest *expected* total payoff for all spreads $s$ at $t$
  **if** $maxVal \neq 0$ **then**
    $s' \leftarrow \text{argmax}_s\, E(\mathbf{\Pi}_t^{tot})$ extract spread hour associated with max *expected* payoff
    $\pi_t^{tot} \leftarrow \pi_t^{tot,(s')}$ extract *actual* total payoff for state $s'$ at $t$
    **if** $\pi_t^{tot} < 0$ **then**
      $n \leftarrow n + 1$
      $L_n \leftarrow \pi_t^{tot}$